\documentclass[seceq]{ptptex}
\usepackage{wrapft}
\usepackage{graphicx}

\usepackage{epsf}

\usepackage{amsmath,amssymb,color,graphics,amscd,amsfonts}

\newcommand{\Tr}{\mathrm{Tr}}

\title{
Phase Structure of 
the Large-$N$ Reduced 
Gauge Theory \\
and 
the Generalized Weingarten Model 
}

\author{
Masanori \textsc{Hanada}$^{1,}$
\footnote{E-mail:hana@gauge.scphys.kyoto-u.ac.jp}, 
Takashi \textsc{Kanai}$^{1,}$
\footnote{E-mail:kanai@gauge.scphys.kyoto-u.ac.jp}, 
Hikaru \textsc{Kawai}$^{1,2,}$
\footnote{E-mail:hkawai@gauge.scphys.kyoto-u.ac.jp},\\
and Fukuichiro \textsc{Kubo}$^{1,}$
\footnote{E-mail:kubo@gauge.scphys.kyoto-u.ac.jp}
}

\inst{
$^1$Department of Physics, Kyoto University, 
Kyoto 606-8502, Japan\\
$^2$ Theoretical Physics Laboratory, RIKEN, Wako 351-0198, 
Japan
}

\recdate{April 9, 2006}

\abst{
We study a generalization of the Weingarten model reduced to a point, 
which becomes the large-$N$ reduced $U(N)$ gauge theory  
in a special limit.
We find that the $U(1)^d$ symmetries 
are broken one by one, 
and subsequently restored
as $U(1)^d\to U(1)^{d-1}\to\cdots\to U(1)\to 1\to U(1)^d$
as we change the coupling constants. 
In this model, we can develop an efficient algorithm,
and we can clearly see the phase structure of the large-$N$ 
reduced model.
We thus conclude that this model would be useful for the study of the
unitary model.
}

\begin{document}

\maketitle

\section{Introduction}
In this paper, we study a model defined by the action
\begin{eqnarray}
  S=
  -\beta N\sum_{\mu\neq\nu}^d \Tr
  \left(A_\mu^\dagger A_\nu^\dagger A_\mu A_\nu\right)
  +
  \kappa N\sum_{\mu=1}^d \Tr
  \left(A_\mu^\dagger A_\mu-1\right)^2,  
  \label{action}
\end{eqnarray}
where $A_\mu$ ($\mu=1,2,\cdots,d$) are complex $N\times N$ matrices.
This model is a generalized version of the reduced Weingarten model.

The original Weingarten model \cite{Weingarten} was proposed
as a nonperturbative description of the Nambu-Goto string.
This model is defined as follows. 
Consider the $d$-dimensional square lattice ${\mathbb Z}^d$ and
introduce a complex $N\times N$ matrix $A_{x,\mu}$
for each link connecting the sites $x$ and $x+\hat{\mu}$
in such a way that $A_{x+\hat{\mu},-\mu}=A_{x,\mu}^\dagger$. 
Then the action of the Weingarten model is given by 
\begin{eqnarray}
  S_W
  =
  -N\gamma\sum_x\sum_{\mu\neq\nu}
  \Tr\left(
    A_{x,\mu}A_{x+\hat{\mu},\nu}
    A^\dagger_{x+\hat{\nu},\mu}A^\dagger_{x,\nu}
  \right)
  +N\sum_x\sum_{\mu=1}^d 
  \Tr\left(
    A_{x,\mu}^\dagger A_{x,\mu}
  \right).
  \label{action:original Weingarten}
\end{eqnarray}
The partition function is given by 
\begin{eqnarray}
  Z_W=\int dm_N \exp\left(-S_W\right), 
\end{eqnarray}
where the measure $dm_N$ is defined by 
\begin{eqnarray}
  dm_N
  =
  \prod_{x,\mu}\prod_{i,j}
  \left(
    \frac{N}{\pi}
    d\left[{\rm Re} (A_{x,\mu})_{ij}\right]
    d\left[{\rm Im} (A_{x,\mu})_{ij}\right]
  \right). 
\end{eqnarray}
Let $C_i$ ($i=1,\cdots,n$) be closed contours on the lattice. 
Multiplying $A_\mu$ along $C_i$ and taking the trace, 
we obtain Wilson loops $w(C_i)$. 
The correlator of $w(C_i)$, defined by 
\begin{eqnarray}
  W(C_1,\cdots,C_n)
  =
  \frac{1}{Z_W}\int dm_N \exp\left(-S_W\right)
  \frac{1}{N}w(C_1)\cdots\frac{1}{N}w(C_n),
  \label{Wilson loop correlator0}
\end{eqnarray}
is evaluated as
\begin{eqnarray}
  W(C_1,\cdots,C_n)
  =
  N^{2-2n}\sum_{s\in S(\{C_i\})}
  \exp\left(
    -a(s)\log\gamma^{-1}-h(s)\log N^2
  \right)
  \label{Wilson loop correlator}
\end{eqnarray}
for large-$N$, where $S(\{C_i\})$ is the set of surfaces on the lattice 
whose boundary is $C_1\cup\cdots\cup C_n$, 
$a(s)$ is the area of the surface $s$, and 
$h(s)$ is the number of handles of $s$. 
If we regard $\log\gamma^{-1}$ and $\frac{1}{N^2}$ 
as the string tension and the string coupling, respectively,  
Eq.(\ref{Wilson loop correlator}) can be interpreted as 
the sum of random surfaces weighted by the Nambu-Goto action. 

Next, let us consider the reduced Weingarten model \cite{EK3}, whose
action is given by
\begin{eqnarray}
  S_{RW}
  =
  -N\gamma\sum_{\mu\neq\nu}
  \Tr\left(
    A_{\mu}A_{\nu}
    A^\dagger_{\mu}A^\dagger_{\nu}
  \right)
  +N\sum_{\mu=1}^d 
  \Tr\left(
    A_{\mu}^\dagger A_{\mu}
  \right). 
  \label{action:reduced Weingarten}
\end{eqnarray}
This action is invariant under the $U(1)^d$ transformation 
\begin{eqnarray}
  A_\mu\to e^{i\theta_\mu}A_\mu. 
\end{eqnarray}
If this symmetry is not broken spontaneously in the large-$N$ limit,
the correlators of the Wilson loops
of this model are equal to those of the
original Weingarten model (\ref{action:original Weingarten}). 
Because the reduced Weingarten model has only $d$ matrices, 
numerical calculations are more tractable.
This model was studied numerically \cite{KO} in the cases $d=2$ and $3$, 
and it was shown that the Weingarten model does not describe 
smooth surfaces, but branched polymers \cite{Kawai}. 

One possibility to overcome this difficulty 
is to consider the action (\ref{action}). 
This action is motivated by the following observation.
In the case of the Hermitian matrix model, 
we can describe a type 0B string 
by flipping the sign of the double-well potential \cite{TT}.
Therefore, we expect that also in the case of the Weingarten model, 
worldsheet supersymmetry is introduced by modifying the potential, 
and it may prevent a worldsheet from 
falling into a branched polymer. 

Although there are several possibilities for modification, 
we choose Eq.(\ref{action}) because 
it has the following interesting properties. 
First, at $\beta=0$, this model consists of 
a set of $d$ copies of a complex one-matrix model with a double-well
potential. 
Secondly, at $\kappa=\infty$, the matrices $A_\mu$ 
are constrained to be unitary, and 
this model becomes the reduced $U(N)$ gauge theory.\cite{EK1} 
This model is worth studying in its own right, 
because of its relation to $U(N)$ gauge theory 
and matrix models of superstring theory.\cite{BFSS,IKKT}  
Thirdly, when $\beta$ and $\kappa$ are finite, this model allows 
a lattice string interpretation similar to that allowed by 
the original Weingarten model, 
because 
the relation (\ref{Wilson loop correlator}) also holds in this model, 
as long as no surface $s$ intersects itself.
Note that this action is the large-$N$ reduction of 
the ``interpolating model'' proposed in Ref.\cite{IMS}, 
in which the second term on the r.h.s. of Eq.(\ref{action:original Weingarten}) 
is replaced with
$
\kappa N\sum_x\sum_{\mu=1}^d 
\Tr\left(
  A_{x,\mu}^\dagger A_{x,\mu}-1
\right)^2$.

This model has been solved analytically only in the special cases
$\beta=0$ \cite{IMS} and $\kappa=\infty$, $d=2$.\cite{PR,GW} 
We investigated the general parameter region numerically
by Monte-Carlo simulation.

The organization of this paper is as follows. 
In \textsection \ref{sec:numerical results} we present 
the numerical results for $d=2$, $3$ and $4$.  
We find $d$ phase transitions that correspond to 
the partial breakdowns of $U(1)^d$ symmetry.
Their $N$ dependences and the existence of the hystereses
indicate that they are of first order.
At $\kappa=\infty$, these phase transitions smoothly approach 
the known phase transitions of the large-$N$ reduced 
$U(N)$ gauge theory.\cite{NN} 
In \textsection \ref{sec:Recovery}
we show that at finite $\kappa$, the $U(1)^d$ symmetry is restored
if we further increase $\beta$.
Because in our model, numerical calculations are more tractable
\footnote{
We have applied the standard metropolis algorithm to 
each element of the $N \times N$ matrices. 
We have stored the matrices $A_{\mu}A_{\nu}$ in addition to 
the original $A_{\mu}$'s. 
In this way, we can reduce the computational time 
for evaluating the variation of the action to order $N$, 
while it is proportional to $N^2$ 
in the case of the unitary matrix model.
For details, see \citen{KO}.
},
it is useful for the study of the unitary theory.
\textsection \ref{sec:Discussion} is devoted to
conclusions and discussion. 

\section{Numerical results}\label{sec:numerical results}
\subsection{Numerical results for $d=4$}

\begin{figure}[bth]
\begin{center}
    \scalebox{1}{\input{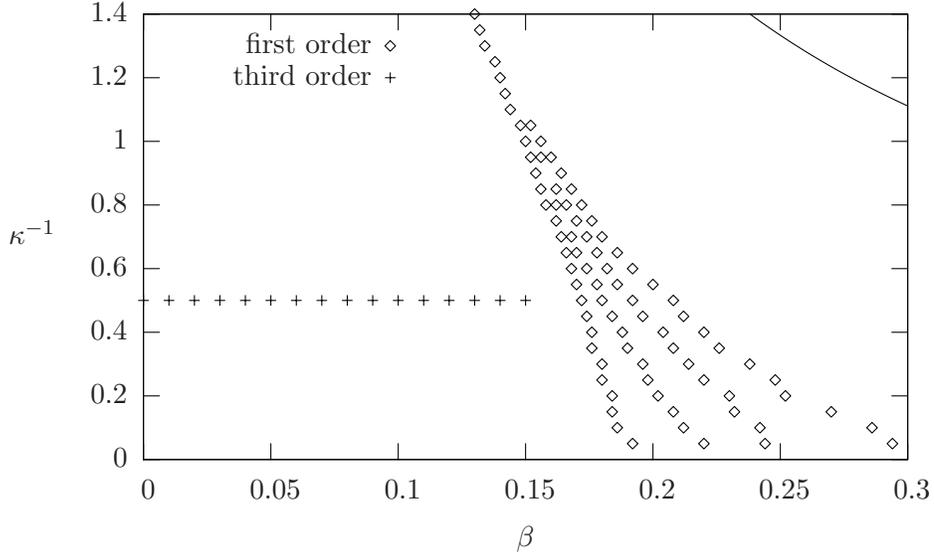}}
  \caption{Phase diagram of the $d=4$ model.
  Phase transitions at points indicated by the symbols 
$\diamond$ seem to be of first order,
  while those indicated by $+$ are of third order.
  Outside the solid line, representing $\beta\kappa^{-1}=\frac{1}{3}$, 
  the action is not bounded from below.   
  }\label{4DphaseTR}
\end{center}
\end{figure}

We begin with the case $d=4$.
The phase diagram here is given by Fig. \ref{4DphaseTR}.
There, the symbols $\Diamond$ represent the points where 
the fluctuation of the action, 
$\langle(\Delta S)^2\rangle
=\langle S^2\rangle-\langle S\rangle^2$, diverges. 
The symbols $+$ represent a third-order phase transition.  
Such a phase structure can be seen clearly only when
$N$ is sufficiently large, $N\gtrsim 30$.

\begin{figure}[bth]
\begin{center}
    \scalebox{1}{\input{onematrix_heat4D-3.tex}}
  \caption{The ``specific heat'' $C$ of the $d=4$ model 
    at $\beta=0, N=35$. The solid curve represents the analytic result. 
  }\label{4Donematrix_heat}
    \scalebox{1}{\input{N16N50.tex}}
    \caption{Plot of $P=\frac{1}{N^2}\sum_{\mu}|\Tr A_{\mu}|^2$ 
      for $N=16$ and $N=50$.
      $C\simeq 2.162$ is a fitting parameter.}\label{N16N50}
  \end{center}
\end{figure}
At $\beta=0$, (\ref{action}) describes
$d$ copies of the complex one-matrix model, whose action is given by 
\begin{eqnarray}
  S_{\mathrm{one-matrix}}=
  \kappa N\Tr
  \left(A^\dagger A-1\right)^2. 
\end{eqnarray}
The free energy is $d$ times as large as that of the complex one-matrix model. 
The second derivative of the free energy with respect to $\kappa$,
$C=\kappa^2\frac{\partial^2}{\partial\kappa^2} (log Z)$,
is given by Ref.~\citen{IMS}: 
\begin{eqnarray}
  \frac{C}{N^2}
  =
  \left\{
    \begin{array}{cc}
      \frac{d}{27}\left(
        \kappa^2+(\kappa-3)\sqrt{\kappa(\kappa+6)}
        +\frac{1}{2} \right)&,\ (\kappa<2)\\
      \frac{d}{2} &.\ (\kappa\ge 2)
    \end{array}
  \right.\label{eq:IMS}
\end{eqnarray}
The derivative of $C$ with respect to $\kappa$ 
is finite but not continuous at $\kappa=2$, indicating a 
Gross-Witten-type phase transition. 
In Fig.~\ref{4Donematrix_heat} we plot $C$ as a function
of $\kappa^{-1}$ for $\beta=0$. 
Our numerical result accurately reproduces the analytic solution.
We found a similar third-order phase transition at
$\kappa^{-1}\simeq 0.5$ with $0\le \beta \lesssim 0.15$.
For $\beta\gtrsim 0.15$, we cannot see this phase transition clearly, 
because it is buried 
in the tail of the peak of the first-order phase transition. 

If we fix $\kappa$ at a sufficiently large value and vary $\beta$,
we find four critical points. 
We call them $\beta_1,\beta_2,\beta_3$ and $\beta_4$ in ascending order. 
They converge to finite values 
as $\kappa\to\infty$. 
For $\kappa^{-1}\gtrsim 1.0$, these phase transitions seem to 
merge. 

Our result is consistent with that found in 
Refs.~\citen{HNT} and \citen{NN} at $\kappa=\infty$. 
In Ref.~\citen{HNT} this parameter region was studied up to $N=16$, and 
it was found that the order parameter 
$P=\frac{1}{N^2}\sum_\mu|\Tr A_\mu|^2$ 
of the $U(1)^4$ symmetry 
behaves asymptotically as $P\sim 4-\frac{C}{\sqrt{2\beta}}$ at large $\beta$, 
where $C\sim 2.162$ is a fitting parameter.  
We plot $P$ in Fig.~\ref{N16N50}. 
It is seen that at $N=16$, our result agrees well with that of Ref.~\citen{HNT}. 
Although the phase transition at $\beta_1$ can be seen clearly,
the phase transitions at $\beta_2,\beta_3$ and $\beta_4$
cannot be seen at this stage.
For $N\gtrsim 30$, we can see them clearly. This is consistent
with the results of Ref.~\citen{NN}.

\begin{figure}[bthp]
  \begin{center}
    \scalebox{1}{\input{4DU1breaking.tex}}
    \caption{$U(1)^4$ breakdown in the unitary matrix model at $N=50$.
      The quantities $\frac{1}{N}|\Tr A_\mu|\ (\mu=1,2,3,4)$ are plotted
      simultaneously.  
    }\label{4DU1breaking}
    \begin{tabular}{lr}    
      \parbox{0.5\textwidth}{
        \scalebox{0.55}{\input{4Dunitary_hyst.tex}}
        \caption{Hysteresis in the phase transition of the unitary
          matrix model around $\beta_1$, $N=50$. 
          $P=\frac{1}{N^2}\sum_{\mu}|\Tr A_{\mu}|^2$.}\label{4Dunitary_hyst}
      }
      \qquad
      \parbox{0.5\textwidth}{
        \scalebox{0.55}{\input{4Dunitary_hyst2.tex}}
        \caption{Hysteresis in the phase transition of the unitary 
          matrix model around $\beta_2$, $N=50$. 
          $P=\frac{1}{N^2}\sum_{\mu}|\Tr A_{\mu}|^2$.
        }\label{4Dunitary_hyst2}
      }
    \end{tabular}
  \end{center}
\end{figure}
In Fig.~\ref{4DU1breaking}, we plot $|\Tr A_\mu|$ simultaneously for
$\mu=1,2,3,4$ at $\kappa=\infty$ and $N=50$. 
For $\beta<\beta_1\simeq 0.190$, we have $|\Tr A_\mu|=0$ for any $\mu$. 
This implies that $U(1)^4$ is not broken in this region. 
For $\beta_1<\beta<\beta_2\simeq 0.219$, the quantity $|\Tr A_\mu|$ becomes
nonzero. This implies that $U(1)^4$ is broken to $U(1)^3$. 
In the same way, for $\beta_2<\beta<\beta_3\simeq 0.257$, 
$|\Tr A_\mu|$ is nonzero for one of the $A_{\mu}$, 
and hence $U(1)^4$ is broken to $U(1)^2$, 
for $\beta_3<\beta<\beta_4\simeq 0.307$ 
$|\Tr A_\mu|$ is nonzero for three $A_{\mu}$ and hence
$U(1)^4$ is broken to $U(1)$,  
and for $\beta>\beta_4$, $|\Tr A_\mu|$ is nonzero for all $A_{\mu}$
and $U(1)^4$ is broken completely.  
Similarly, for finite $\kappa$, 
$U(1)^4$ breaks to $U(1)^3$ at $\beta_1$, to $U(1)^2$ at $\beta_2$, 
to $U(1)$ at $\beta_3$, and is broken completely at $\beta_4$. 

In Fig.~\ref{4Dunitary_hyst} and Fig.~\ref{4Dunitary_hyst2}, 
we plot 
$P=\frac{1}{N^2}\sum_\mu|\Tr A_\mu|^2$ versus $\beta$ for 
$\kappa=\infty$. 
Because there are hystereses around $\beta_1$ and $\beta_2$,  
these phase transitions are of first order. 
We conjecture that the other transitions are also of first order, 
although we have not seen yet clear hysteresis.  
The phase transitions at finite $\kappa$ also seem to be of first
order. 

\subsection{Numerical result for $d=3$}
\hspace{0.51cm}
In this subsection, we consider the case $d=3$.
As we can see in Fig.~\ref{3DphaseTR}, 
at large, fixed $\kappa$ there are three curves of first-order phase
transitions.  
\begin{figure}[tbhp]
  \begin{center}
      \scalebox{1}{\input{3DphaseTR-1.tex}}
      \caption{Phase diagram of the $d=3$ model.
      Outside the solid curve, the action $\beta\kappa^{-1}=\frac{1}{2}$
      is not bounded from below. }\label{3DphaseTR}
  \end{center}
  \begin{center}
    \scalebox{1}{\input{3DU1breaking.tex}}
    \caption{$U(1)^3$ breakdown in the unitary matrix model at $N=50$.
      $\frac{1}{N}|\Tr A_\mu|\ (\mu=1,2,3)$ are plotted
      simultaneously.  
    }\label{3DU1breaking}
\end{center}
\end{figure}
We call them $\beta_1,\beta_2$ and $\beta_3$  in ascending order. 
$U(1)^3$ is broken to $U(1)^2$ at $\beta_1$, to $U(1)$ at $\beta_2$, 
and completely at $\beta_3$. 
The quantities $\beta_1,\beta_2$ and $\beta_3$ converge to finite values 
as $\kappa^{-1}\to 0$. 

In Fig.~\ref{3DU1breaking}  
we plot $|\Tr A_\mu|$ simultaneously for $\mu=1,2,3$ at $\kappa^{-1}=0$. 
For $\beta<\beta_1\simeq 0.30$, $|\Tr A_\mu|=0$ for all $\mu$. 
This implies that $U(1)^3$ is not broken in this region. 
For $\beta_1<\beta<\beta_2\simeq 0.40$, $|\Tr A_\mu|$ becomes nonzero 
for one $A_{\mu}$. This implies that $U(1)^3$ is broken to $U(1)^2$.  
In the same way, for $\beta_2<\beta<\beta_3\simeq 0.57$, 
$|\Tr A_\mu|$ is nonzero for two $A_{\mu}$, and hence
$U(1)^3$ is broken to $U(1)$, 
and for $\beta>\beta_3$
$|\Tr A_\mu|$ is nonzero for all $A_{\mu}$, 
and $U(1)^3$ is broken completely.  
For $\kappa^{-1}\gtrsim 1.5$, the three curves of the phase transitions 
seem to merge. 

\subsection{Numerical results of $d=2$}

\begin{figure}[btph]
  \begin{center}
    \scalebox{1}{\input{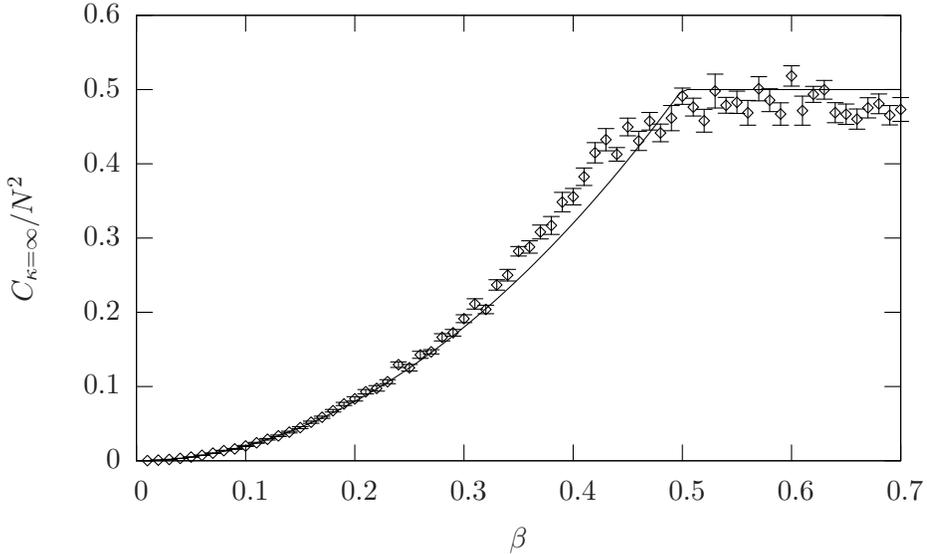}}
    \caption{The specific heat of the $d=2$ model 
      at $\kappa^{-1}=0, N=25$. 
      The solid curve is the analytic result (\ref{eq:Gross Witten}). } 
    \label{2Dunitary_heat}
  \end{center}
\end{figure}

In this subsection we consider the case $d=2$.
In this case, if $\kappa^{-1}=0$,
this model is equivalent to the Gross-Witten model,
and the specific heat $C=\beta^2\frac{\partial^2}{\partial\beta^2}Z$ 
is given by \citen{PR,GW}
\begin{eqnarray}
  \frac{C}{N^2}
  =
  \left\{
    \begin{array}{cc}
      2\beta^2 &,~ \left(\beta\le\frac{1}{2}\right)\\
      \frac{1}{2}&.~ \left(\beta>\frac{1}{2}\right)
    \end{array}
  \right
  .\label{eq:Gross Witten}
\end{eqnarray}
A third-order phase transition takes place at $\beta=\frac{1}{2}$. 
This can also be observed by numerical analysis 
(see Fig.~\ref{2Dunitary_heat}).  

\begin{figure}[btph]
  \begin{center}
      \scalebox{1}{\input{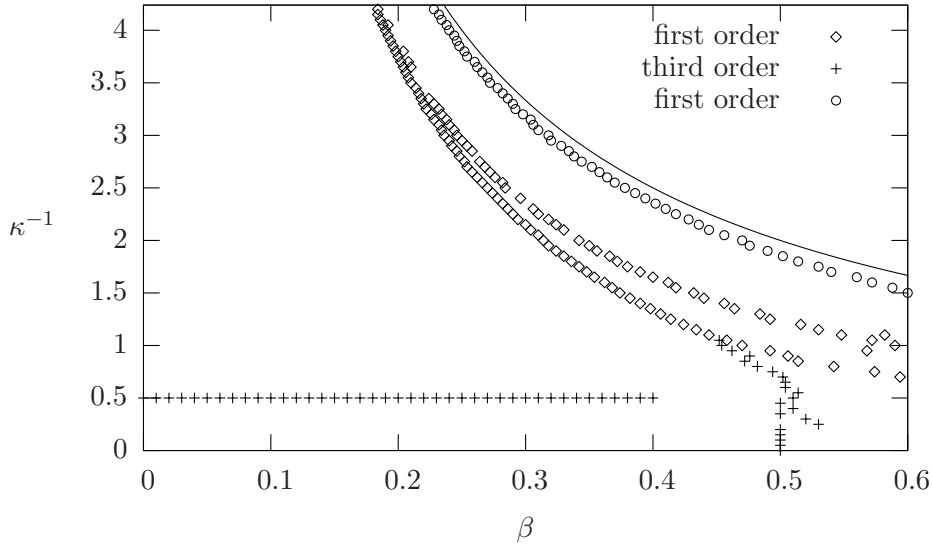}}
      \caption{Phase diagram of the $d=2$ model.
        Outside the solid curve, representing $\beta\kappa^{-1}=1$,
        the action is not bounded from
        below. 
      }\label{2DphaseTR}
  \end{center}
\end{figure}

The phase diagram for $N=50$ is given in Fig.~\ref{2DphaseTR}.\footnote{
For $\kappa=2$ and $0.25\le\beta<0.4$, we studied up to $N=75$ and found 
no other phase transition.  
}
At large, fixed $\kappa$ there are two curves of first-order phase
transitions.
We call them $\beta_1$ and $\beta_2$ in ascending order. 
They correspond to the breakdown of $U(1)^2$ symmetry. 
If we increase $\beta$ with $\kappa$ fixed, 
first $U(1)^2$ is broken to $U(1)$ at $\beta_1$, 
and it is broken completely at $\beta_2$. 

At small $\kappa$, $\beta_1$ and $\beta_2$ seem to merge. 
They seem to go to $\infty$ as $\kappa^{-1}\to 0$. 
This is consistent with the analytic result, 
in which $U(1)^2$ is not broken.

In Fig.~\ref{2DphaseTR} there is a line of the symbols $\circ$.
Beyond this line, the $U(1)^2$ symmetry is restored.
We discuss in detail this novel phenomena in the next subsection.

\subsection{Restoration of $U(1)^d$ symmetry}\label{sec:Recovery}
\hspace{0.51cm}
In this subsection we show that the $U(1)^d$ symmetry
is restored as $\kappa^{-1}$ or $\beta$ becomes larger.
This restoration takes place in the well-defined region
where the action is bounded from below. 

\begin{figure}[bth]
  \begin{center}
    \input{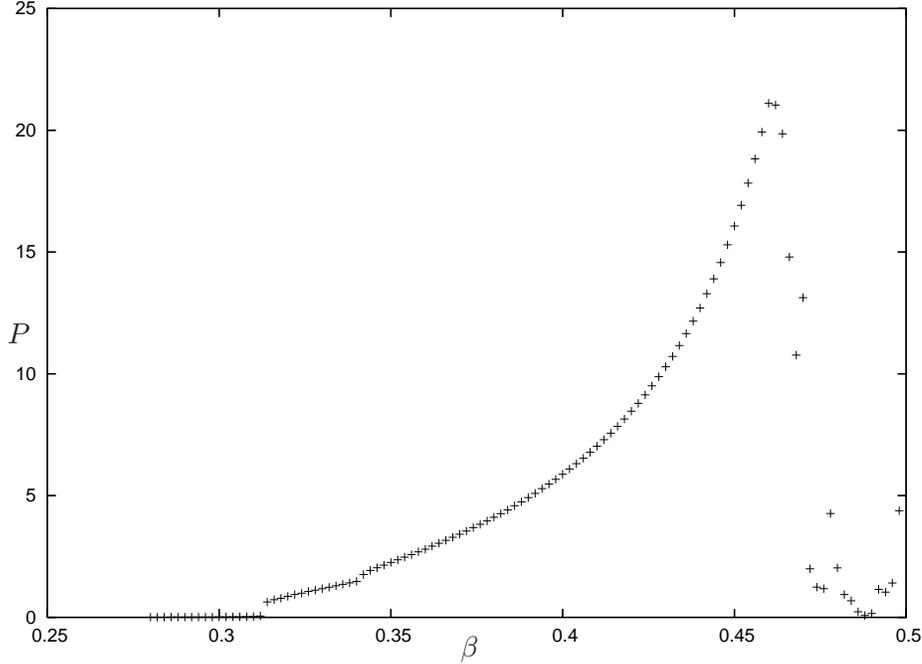}
    \caption{Plot of  $P=\frac{1}{N^2}\sum_{\mu}|\Tr A_{\mu}|^2$
      for $\kappa^{-1}=2.0$ and $N=50$. $U(1)^2$ is broken at 
      $\beta\simeq 0.315$ and $\beta\simeq 0.340$, and then
      restored at $\beta\simeq 0.48$.
    }\label{N50IK2}
  \end{center}
\end{figure}

In order to see whether the action is bounded from below,
it is enough to consider the quartic term
\begin{eqnarray}
  S|_{\mathrm{quartic}}=
  -\beta N\sum_{\mu\neq\nu}^d \Tr
  \left(A_\mu^\dagger A_\nu^\dagger A_\mu A_\nu\right)
  +
  \kappa N\sum_{\mu=1}^d \Tr
  \left(A_\mu^\dagger A_\mu A_\mu^\dagger A_\mu\right).  
  \label{quartic}
\end{eqnarray}
This is positive definite if and only if \footnote{
This condition comes from the following steps. 
First, the inequality of the arithmetic and geometric means, 
\begin{eqnarray*}
  2\mathrm{Re}\  \Tr\left( A B^\dagger\right) \le \Tr \left(A A^\dagger\right) 
  + \Tr \left(B B^\dagger\right) ,
\end{eqnarray*}
implies
\begin{eqnarray*}
  2\mathrm{Re}\  \Tr\left( A_\mu A_\nu A_\mu^\dagger A_\nu^\dagger
  \right) 
  &\le& \Tr \left(A_\mu^\dagger A_\mu A_\nu A_\nu^\dagger\right)
          +\left(A_\mu A_\mu^\dagger A_\nu^\dagger A_\nu\right) \\
  &\le& \Tr \left(A_\mu^\dagger A_\mu A_\mu^\dagger A_\mu\right)
          +\left(A_\nu^\dagger A_\nu A_\nu^\dagger A_\nu\right).
\end{eqnarray*}
Thus, summing the spacetime subscripts, we get
\begin{eqnarray*}
  \sum_{\mu\neq\nu}^d \Tr
  \left(A_\mu^\dagger A_\nu^\dagger A_\mu A_\nu\right)
  \le
  (d-1)\sum_{\mu=1}^d \Tr
  \left(A_\mu^\dagger A_\mu \right)^2.
\end{eqnarray*}
Equality here is realized, e.g., in the case of the unit matrix.
}
$\beta \kappa^{-1} \le \frac{1}{d-1}$.
In Fig.~\ref{N50IK2}, we plot $P=\frac{1}{N^2}\sum_\mu|\Tr A_\mu|^2$
at $\kappa^{-1}=2.0$.
For $\beta \simeq 0.315$ and $0.340$, there are discontinuities
in the value of $P$. These correspond to the breakdowns of 
$U(1)^2$ symmetry.
As we approach the boundary of the well-defined region, $\beta=0.5$, 
the value of $P$ decreases to zero, 
which suggests that the $U(1)^2$ symmetry is restored.
In Fig.\ref{2DphaseTR}, we denote this phase transition
by the symbols $\circ$.

Although the signal is not clear, we conjecture that 
the restoration of $U(1)^d$ takes place also for $d=3$ and $4$.
Because such a phase transition would take place 
near the boundary of the well-defined region,
we believe that it does not occur in the unitary model.

\section{Conclusions and discussion}\label{sec:Discussion}

We have found $d$ first-order phase transitions
corresponding to one-by-one breakdowns of the $U(1)^d$ symmetry.
As $\kappa^{-1}\to 0$, they are smoothly connected to
those found in Ref.~\citen{NN}.
We also found a first-order phase transition
corresponding to the restoration of $U(1)^d$.
These phase transitions can be clearly seen 
for $N\gtrsim 30$. 
In the region where $c$ $(\le d)$ of the $U(1)$ are not broken, 
our model can be regarded as a 
model defined on a $c$-dimensional lattice.
In particularly, at $\kappa^{-1}=0$, it may represent a 
$c$-dimensional Yang-Mills theory coupled to
$(d-c)$ matter fields.

Although the first-order phase transitions mentioned above
may not allow a continuum limit, 
there is a continuous curve of third-order phase transitions which
may be equivalent to a lattice string theory.
In order to determine whether it has a continuum limit, 
we must study the expectation values of Wilson loops and the Creutz ratio.
It would also be interesting to study the ``interpolating model'',\cite{IMS}  
which does not reduce to a point.  
Because our reduced model (\ref{action}) and the interpolating model
are not equivalent for $\beta>\beta_1$, 
where $U(1)^d$ is broken, there is still a possibility that 
the interpolating model describes a continuum string in this
region. 

In the large-$N$ $U(N)$ gauge theory on a lattice, 
there is a bulk phase transition, 
which does not involve the breakdown of any symmetry.  
In our simulation, however, we do not find a corresponding 
phase transition. However, this is not a contradiction, because 
in the large-$N$ reduced $U(N)$ gauge theory, 
the bulk phase transition is hidden by the breakdown 
of $U(1)^d$ symmetry \cite{NN}. 
We hypothesize that a similar phenomenon occurs also for $\kappa<\infty$. 

The large-$N$ reduced $U(N)$ gauge theory
can be regarded as a toroidal compactification of 
the bosonic part of IIB matrix model.\cite{IKKT} 
In this interpretation, the $U(1)^d$ symmetry bocomes the 
translational symmetry, 
and there emerge spacetimes with various dimensions, 
depending on the coupling constant.

\section*{Acknowledgements}

Numerical computations in this work was carried out at
Yukawa Institute Computer Facility and 
RIKEN Super Combined Cluster System. 
The authors thank Tatsumi Aoyama, Yuri Makeenko, Yoshinori Matsuo 
and Jun Nishimura for stimulating discussions and comments. 
M. H. would like to
thank the Japan Society for the Promotion of Science for financial
support. 
This work was also supported in part by a Grant-in-Aid for
the 21st Century COE ``Center for Diversity and Universality in
Physics''.


\begin{thebibliography}{99}

\bibitem{Weingarten}
D. Weingarten, 
Phys. Lett. B~\textbf{90} (1980), 280. 

\bibitem{EK3}
T. Eguchi and H. Kawai, 
Phys. Lett. B~\textbf{114} (1982), 247. 

\bibitem{KO}
H. Kawai and Y. Okamoto,
Phys. Lett. B~\textbf{130} (1983), 415. 

\bibitem{Kawai}
H. Kawai, 
Nucl. Phys. B (Proc. Suppl.) \textbf{26} (1992), 93. 

\bibitem{TT}
T. Takayanagi and N. Toumbas, 
J. High Energy Phys. 07 (2003), 064; 
hep-th/0307083.\\ 
M. R. Douglas, I. R. Klebanov, D. Kutasov, J. Maldacena, E. Martinec
and N. Seiberg, 
{\it A new hat for the c = 1 matrix model},
in Shifman, M. (ed.) et al.: From fields to strings, vol. 3:1758; 
hep-th/0307195. 

\bibitem{EK1}
T. Eguchi and  H. Kawai, 
Phys. Rev. Lett. \textbf{48} (1982), 1063. 

\bibitem{BFSS}
T. Banks, W. Fischler, S. Shenker and L. Susskind, 
Phys. Rev. D~\textbf{55} (1997), 5112; hep-th/9610043.

\bibitem{IKKT}
N. Ishibashi, H. Kawai, Y. Kitazawa and A. Tsuchiya, 
Nucl. Phys. B~\textbf{498} (1997), 467; hep-th/9612115.

\bibitem{IMS}
E.-M. Ilgenfritz, Y. M. Makeenko and T. V. Shakhbazian, 
Phys. Lett. B~\textbf{172} (1986), 81. 

\bibitem{PR}
G. Paffuti and P. Rossi, 
Phys. Lett. B~\textbf{92} (1980), 321. 

\bibitem{GW}
D. J. Gross and E. Witten, 
Phys. Rev. D~\textbf{21} (1980), 446. 

\bibitem{HNT}
T. Hotta, J. Nishimura and A. Tsuchiya, 
Nucl. Phys. B~\textbf{545} (1999), 543; 
hep-th/9811220.

\bibitem{NN}
  R. Narayanan and H. Neuberger,
  Phys. Rev. Lett. \textbf{91} (2003), 081601; 
  hep-lat/0303023. \\
  J. Kiskis, R. Narayanan and H. Neuberger,
  Phys. Lett. B~\textbf{574} (2003), 65; 
  hep-lat/0308033.

\end{thebibliography}
\end{document}